\documentclass{ws-ijmpcs}

\begin{document}

\markboth{Guido Zavattini}
{Measuring the magnetic birefringence of vacuum: the PVLAS experiment}

%
\catchline{}{}{}{}{}
%

\title{MEASURING THE MAGNETIC BIREFRINGENCE OF VACUUM:\\
THE PVLAS EXPERIMENT
}

\author{G. ZAVATTINI\footnote{
Corresponding author: Guido Zavattini,  INFN - sezione di Ferrara and Physics Department, University of Ferrara, Via Saragat 1, Blocco C, 44122 Ferrara, Italy. e-mail: zavattini@fe.infn.it}}
\address{INFN - Sez. di Ferrara and Dip. di Fisica, Universit\`a di Ferrara\\ via Saragat 1, Blocco C, I-44122 Ferrara, Italy}

\author{U. GASTALDI, R. PENGO, G. RUOSO}
\address{INFN - Lab. Naz. di Legnaro\\ viale dell'Universit\`a 2, I-35020 Legnaro, Italy}

\author{F. DELLA VALLE, E. MILOTTI}
\address{INFN - Sez. di Trieste and Dip. di Fisica, Universit\`a di Trieste\\ via A. Valerio 2, I-34127 Trieste, Italy}

\maketitle


\begin{abstract}
We describe the principle and the status of the PVLAS experiment which is presently running at the INFN section of Ferrara, Italy, to detect the magnetic birefringence of vacuum. This is related to the QED vacuum structure and can be detected by measuring the ellipticity acquired by a linearly polarized light beam propagating through a strong magnetic field. Such an effect is predicted by the Euler-Heisenberg Lagrangian. The method is also sensitive to other hypothetical physical effects such as axion-like particles and in general to any fermion/boson millicharged particle.
Here we report on the construction of our apparatus based on a high finesse ($>2\cdot10^5$) Fabry-Perot cavity and two 0.9 m long 2.5 T permanent dipole rotating magnets, and on the measurements performed on a scaled down test setup. With the test setup we have improved by about a factor 2 the limit on the parameter $A_e$ describing non linear electrodynamic effects in vacuum: $A_e < 2.9\cdot10^{-21}$ T$^{-2}$ @ 95\% c.l.

\end{abstract}

\ccode{PACS numbers: 11.25.Hf, 123.1K}

\section{Physics case}	

Magnetic vacuum birefringence or light-light interaction in vacuum at very low energies have yet to be observed. Several experimental efforts are underway\cite{HypIn}\cdash\cite{heinzl}
to detect such effects. Indeed QED predicts non linear effects leading to birefringence and light-light scattering (LbL) through the box diagram.\cite{QED}\cdash\cite{BernardOld} Furthermore hypothetical ideas such as the existence of axion-like particles (ALPs) coupling to two photons\cite{Petronzio}\cdash\cite{Raffelt} or the existence of fermion/boson millicharged particles (MCPs)\cite{Gies}\cdash\cite{spin0} could generate both magnetic birefringence and magnetic dichroism. Finally the coupling of four photons through $q\mathchar'26\mkern-10muq$ fluctuations is also possibile but an evaluation of such a contribution to the vacuum magnetic birefringence cannot be extracted from indirect measurements.\cite{qqbar1,qqbar2} Due to the quark masses, though, this last contribution can be expected to be very small.

\begin{figure}[pb].
\centerline{\psfig{file = 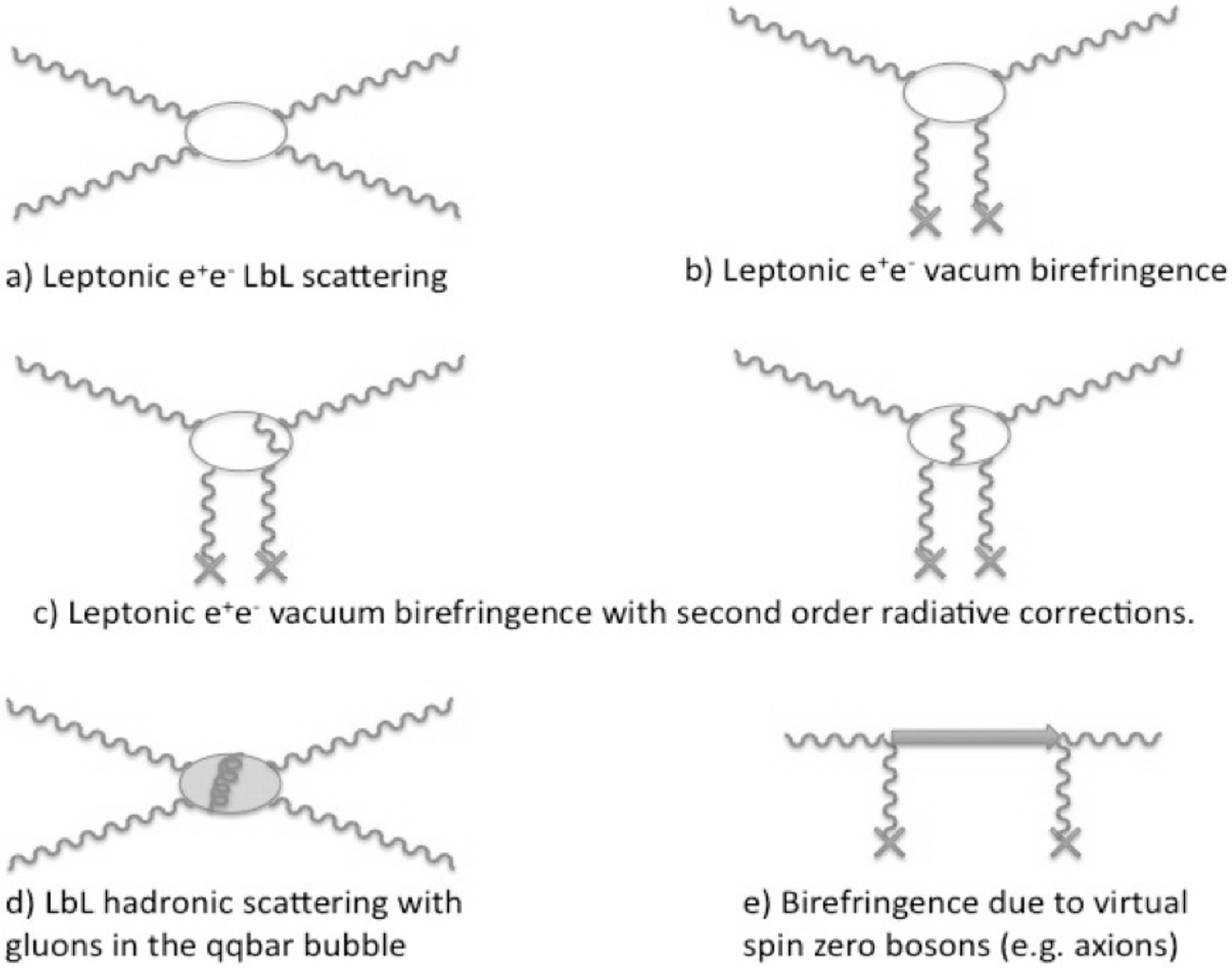, angle = -90, width = 10 cm}}
\caption{Feynman diagrams for four field interactions.}
\label{4photon}
\end{figure}

These different contributions to the four photon interaction are summarized in Figure \ref{4photon}.

Today the best limit on four photon interactions has been set by the PVLAS collaboration\cite{scatter} with an upper bound on vacuum magnetic birefringence $\Delta n^{\rm (PVLAS)} \; @\; 2.3 \;{\rm T}$
\begin{equation}
\Delta n^{\rm (PVLAS)} < 1.0\cdot10^{-19}\; @\; 1064 \; {\text {nm and 2.3 T}}
\end{equation}
which, translated into light-light elastic scattering,\cite{DeTollis}\cdash\cite{BernardOld} results in an upper bound on the
cross section $\sigma_{\gamma \gamma}^{\rm (PVLAS)}$
\begin{equation}
\sigma_{\gamma \gamma}^{\rm (PVLAS)} < 4.6\cdot10^{-58}\; {\text {cm}}^{2} \;@\; 1064 \; {\text {nm}}
\end{equation}
The predicted QED value of the magnetic vacuum birefringence (see below) $\Delta n^{\rm (QED)}$ and light-light elastic scattering cross section $\sigma_{\gamma \gamma}^{\rm (QED)}$ are
\begin{eqnarray}
\label{Deltan}
\Delta n^{\rm (QED)}& =& 2.1\cdot10^{-23}\; @\; 2.3 \; {\text {T}}\\
\sigma_{\gamma \gamma}^{\rm (QED)}& =& 1.8\cdot10^{-65}\; {\text {cm}}^{2} \;@\; 1064 \; {\text {nm}}
\end{eqnarray}

\subsection{Electrodynamics}
In the absence of matter, Maxwell's equations can be obtained from the classical electromagnetic Lagrangian density ${\cal L}_{\rm Cl}$ (in S.I. units)
\begin{equation}
{\cal L}_{\rm Cl} = \frac{1}{2\mu_{\rm 0}}\left(\frac{E^{2}}{c^{2}}-B^{2}\right)
\end{equation}
It is well known that in this case the superposition principle holds thereby excluding light-light scattering and other non linear electromagnetic effects in vacuum.

With the introduction of Dirac's equation for electrons and Heisenberg's Uncertainty Principle, Euler and Heisenberg in 1936\cite{QED} derived a Lagrangian density which leads to electromagnetic non linear effects even in vacuum. For photon energies well below the electron mass and fields much smaller than their critical values, $B \ll B_{\rm crit} = {m_{e}^{2}c^{2}}/{e \hbar} = 4.4\cdot10^{9}$ T, $E \ll E_{\rm crit} = {m_{e}^{2}c^{3}}/{e \hbar} = 1.3\cdot10^{18}$~V/m, the Euler-Heisenberg Lagrangian correction can be written as
\begin{equation}
{\cal L}_{\rm EH} = \frac{A_{e}}{\mu_{\rm 0}}\Bigg[\Big(\frac{E^{2}}{c^{2}}-B^{2}\Big)^{2}+7\Big(\frac{\vec{E}}{c}\cdot\vec{B}\Big)^{2}\Bigg]
\end{equation}
where $\mu_{0}$ is the magnetic permeability of vacuum and
\begin{equation}
A_{e} = \frac{2}{45\mu_{0}}\frac{\alpha^{2}\mathchar'26\mkern-10mu\lambda_e^{3}}{m_{e}c^{2}} = 1.32\cdot10^{-24} \;{\text T}^{-2}
\label{Ae}
\end{equation}
with $\mathchar'26\mkern-10mu\lambda_e$ being the Compton wavelength of the electron, $\alpha={e^2}/{(\hbar c 4\pi\epsilon_0)}$ the fine structure constant, $m_e$ the electron mass, $c$ the speed of light in vacuum. 

This Lagrangian correction allows four field interactions and can be represented, to first order, by the Feynman diagrams shown in Figure \ref{4photon} a) and b).
 Figure \ref{4photon} a) represents light by light scattering whereas Figure \ref{4photon} b) represents the interactions of real photons with a classical field leading to vacuum magnetic birefringence.
 
 To determine the magnetic birefringence of vacuum one can proceed by determining the electric displacement vector $\vec D$ and magnetic intensity vector $\vec H$ from the total Lagrangian density $\cal L = \cal L_{\rm Cl} + \cal L_{\rm EH}$ by using the constitutive relations\cite{Adler}
\begin{equation}
 \vec D = \frac{\partial {\cal L}}{\partial \vec E} \quad {\rm and} \quad \vec H = - \frac{\partial {\cal L}}{\partial \vec B}
 \end{equation}
 From these one obtains
 \begin{eqnarray}
 \vec D &=&\epsilon_{\rm 0}\vec E + \epsilon_{0}A_{e}\Big[4\Big(\frac{E^{2}}{c^{2}}-B^{2}\Big)\vec E + 14 \Big(\vec E\cdot\vec B\Big)\vec B\Big] \label{D}
\\
 \vec H &=& \frac{\vec B}{\mu_{\rm 0}} + \frac{A_{e}}{\mu_{\rm 0}}\Big[4\Big(\frac{E^{2}}{c^{2}}-B^{2}\Big)\vec B - 14 \Big(\frac{\vec E\cdot\vec B}{c^2}\Big)\vec E\Big]
 \label{H}
 \end{eqnarray} 
 With the $\vec D$ and $\vec H$ vectors one can use Maxwell's equation in media to now describe light propagation in an external field. It is evident that these will no longer be linear due to the non linear dependence of $\vec D$ and $\vec H$ with respect to $\vec E$ and $\vec B$. 
 By assuming a linearly polarized beam of light propagating perpendicularly to an external magnetic field $\vec B_{\rm ext}$ there are two possible configurations: light polarization parallel to $\vec B_{\rm ext}$ and light polarization perpendicular to $\vec B_{\rm ext}$. By substituting $\vec E = \vec E_{\gamma}$ and $\vec B = \vec B_{\gamma} + \vec B_{\rm ext}$ in (\ref{D}) and (\ref{H}) one finds the following relations for the relative dielectric constants and magnetic permeabilities
 \begin{eqnarray}
 \left\{ \begin{array}{ll}
 \epsilon_{\parallel} &= 1 + 10 A_{e}B_{\rm ext}^2\\
 \mu_{\parallel} &= 1 + 4 A_{e}B_{\rm ext}^2\\
n_{\parallel} &= 1 + 7 A_{e}B_{\rm ext}^2
\end{array} \right.
\quad
 \left\{ \begin{array}{ll}
 \epsilon_{\perp} &= 1 - 4 A_{e}B_{\rm ext}^2\\
 \mu_{\perp} &= 1 + 12 A_{e}B_{\rm ext}^2\\
n_{\perp} &= 1 + 4 A_{e}B_{\rm ext}^2
\end{array} \right.
\label{index}
 \end{eqnarray}

From these sets of equations two important consequences are apparent: the velocity of light in the presence of an external magnetic field is no longer $c$ and vacuum is birefringent with
 \begin{equation}
 \Delta n = 3 A_{e} B_{\rm ext}^2
 \label{birifQED}
 \end{equation}
 Numerically this leads to the value given in equation (\ref{Deltan}).
\subsection{Post-Maxwellian generalization}
It is interesting to generalize the non linear electrodynamic Lagrangian density correction by introducing three free parameters $\xi$, $\eta_{\rm 1}$ and $\eta_{\rm 2}$:
\begin{equation}
{\cal L}_{\rm pM} = \frac{\xi}{2 \mu_{\rm 0}}\left[\eta_{\rm 1}\left(\frac{E^{2}}{c^{2}}-B^{2}\right)^{2}+4\eta_{\rm 2}\left(\frac{\vec{E}}{c}\cdot\vec{B}\right)^{2}\right]
\end{equation}
where $\xi = 1/B_{\rm crit}^{2} = \left(\frac{e \hbar}{m_{e}^{2} c^{2}}\right)^{2}$ whereas $\eta_{\rm 1}$ and $\eta_{\rm 2}$ are dimensionless parameters depending on the model. With such a formulation the birefringence induced by an external magnetic field is
 \begin{equation}
 \Delta n = 2 \xi\left(\eta_{\rm 2}-\eta_{\rm 1}\right) B_{\rm ext}^2
 \end{equation}
This expression reduces to (\ref{Deltan}) if $\eta_{\rm 1} = \alpha /45 \pi$ and $\eta_{\rm 2} = 7/4 \eta_{\rm 1}$. It is thus apparent that $n_{\parallel}$ depends only on $\eta_{1}$ whereas $n_{\perp}$ depends only on $\eta_{2}$. It is also noteworthy that if $\eta_{\rm 1} = \eta_{\rm 2}$, as is the case in the Born-Infeld model, then there is no magnetically induced birefringence even though elastic scattering will be present.\cite{Haissinsky,Denisov}
 \subsection{New physics}
\label{sec:3}
As mentioned above, two other important hypothetical effects could also cause $n \ne 1$ in the presence of an external magnetic (or electric) field transverse to the light propagation direction. These can be due either to neutral bosons weakly coupling to two photons called axion-like particles (ALP),\cite{Petronzio}\cdash\cite{Raffelt} or millicharged particles (MCP).\cite{Gies}\cdash\cite{spin0} In this second case both fermions and spin-0 particles can be treated.
\subsubsection{ALP}
Search for axions using laboratory optical techniques was experimentally pioneered by the BFRT collaboration\cite{Cameron} and subsequently continued by the PVLAS effort.\cite{HypIn,PRD,scatter} Initially, this second experiment published the detection of a dichroism induced by the magnetic field\cite{PRL} in vacuum. Such a result, although in contrast with the CAST experiment,\cite{CAST} could have been due to axion-like particles. Subsequently the result was excluded by the same collaboration\cite{PRD,scatter} after a series of upgrades to their apparatus and almost simultaneously the axion-like interpretation was excluded by two groups\cite{RizzoALP1,RizzoALP2,gammeV} in a regeneration type measurement. However, the original publication revived interest in the optical effects which could be caused by ALP's and later MCP's.

The Lagrangian densities describing the interaction of either pseudoscalar fields $\phi_{\rm a}$ or scalar fields $\phi_{\rm s}$ with two photons can be expressed as (for convenience, written in natural Heavyside-Lorentz units)
\begin{equation}
{\cal L}_{a} = \frac{1}{M_{a}}\phi_{a} \vec{E}\cdot \vec{B} \quad{\rm and}\quad
{\cal L}_{s} =  \frac{1}{M_{s}}\phi_{s} \left(E^{2}-B^{2} \right)
\label{lagalp}
\end{equation}
where $M_{a}$ and $M_{s}$ are the coupling constants. 

Therefore in the presence of an external uniform magnetic field $\vec{B}_{\rm ext}$ a photon with electric field $\vec{E}_{\rm \gamma}$ parallel to $\vec{B}_{\rm ext}$ will interact with the pseudoscalar field whereas for electric fields perpendicular to $\vec{B}_{\rm ext}$ no such interaction will exist. For the scalar case the opposite is true: an interaction will exist if $\vec{E}_{\rm \gamma} \perp \vec{B}_{\rm ext}$ and will not if $\vec{E}_{\rm \gamma} \parallel \vec{B}_{\rm ext}$. When an interaction is present, an oscillation between the photon and the pseudoscalar/scalar field will exist.

For photon energies above the mass $m_{\rm a,s}$ of such particle candidates, a real production can follow. This will cause an oscillation of those photons whose polarization allows an interaction into such particles. On the other hand, even if the photon energy is smaller than the particle mass, virtual production will follow and will cause a phase delay for those photons with an electric field direction allowing an interaction.

The attenuation $\kappa$ and phase delay $\phi$ for light with polarization allowing an interaction can be expressed, in both the scalar and pseudoscalar cases, as:\cite{Petronzio,Sikivie,Cameron}

\begin{equation}
\kappa =2\left(\frac{B_{\rm ext}L}{4M_{a,s}}\right)^{2}\left(\frac{\sin x}{x}\right)^{2}\quad{\rm and}\quad
\phi = \frac{\omega B_{\rm ext}^{2}L}{2M_{a,s}^{2}m_{a,s}^{2}}\left(1-\frac{\sin2x}{2x}\right)
\end{equation}
where, in vacuum, $x=\frac{Lm_{a,s}^{2}}{4\omega}$, $\omega$ is the photon energy and $L$ is the magnetic field length. The above expressions are in natural Heavyside-Lorentz units whereby 1~T $=\sqrt{\frac{\hbar^{3}c^{3}}{e^{4}\mu_{0}}}= 195$~eV$^2$ and 1~m $=\frac{e}{\hbar c}=5.06\cdot10^{6}$~eV$^{-1}$.  The phase delay $\phi$ is related to the index of refraction $n$ by
\begin{equation}
\phi=k\left(n-1\right)L
\end{equation}
Therefore in the pseudoscalar case, where $n^{a}_{\parallel}>1$ and $n^{a}_{\perp}=1$, and in the scalar case, where $n^{s}_{\perp}>1$ and $n^{s}_{\parallel}=1$, the birefringence $\Delta n = n_{\parallel} - n_{\perp}$ will be
\begin{equation}
\Delta n = n_{\parallel}^{a}-1= 1-n_{\perp}^{s}= \frac{B_{\rm ext}^{2}}{2M_{a,s}^{2}m_{a,s}^{2}}\left(1-\frac{\sin2x}{2x}\right)
\label{pseudo}
\end{equation}

In the approximation for which $x\ll1$ (small masses) this expression becomes
\begin{equation}
\Delta n = n^{a}_{\parallel}-1 = 1-n^{s}_{\perp} = \frac{B_{\rm ext}^{2}m_{a,s}^{2}L^{2}}{16M_{a,s}^{2}}
\end{equation}
whereas for $x \gg 1$
\begin{equation}
\Delta n = n^{a}_{\parallel}-1 = 1-n^{s}_{\perp} = \frac{B_{\rm ext}^{2}}{2M_{a,s}^{2}m_{a,s}^{2}}
\end{equation}
The different behavior of $n^{s}_{\perp}-1$ and $n^{a}_{\parallel}-1$ with respect to $L$ in the two cases where $x\ll1$ and $x\gg1$ is interesting and leaves, in principle, a free experimental handle for distinguishing between various scenarios.

\subsubsection{MCP}
Consider now the vacuum fluctuations of particles with charge $\pm\epsilon e$ and mass $m_{\epsilon}$ as discussed by Gies and Ringwald in references \refcite{Gies} and \refcite{Ringwald}. The photons traversing a uniform magnetic field may interact with such fluctuations resulting in both a pair production if the photon energy $\omega > 2m_{\epsilon}$ and only a phase delay if $\omega < 2m_{\epsilon}$. Furthermore, either fermions or spin-0 charged bosons could exist. Since we are discussing birefringence effects only the (real) index of refraction will be considered here.

 - Dirac fermions

Let us first consider the case in which the millicharged particles are Dirac fermions (Df). As derived by Tsai in reference \refcite{fermion} the indices of refraction of photons with polarization respectively parallel and perpendicular to the external magnetic field have two different mass regimes defined by a dimensionless parameter $\chi$ (S.I. units):
\begin{equation}
\chi\equiv
\frac{3}{2}\frac{\hbar\omega}{m_{\epsilon}c^{2}}\frac{\epsilon e B_{\rm ext}\hbar}{m_{\epsilon}^{2}c^{2}}
\label{chi}
\end{equation}
It can be shown that\cite{Gies,landaulev}
\begin{equation}
n_{\parallel,\perp}^{Df}=1+I_{\parallel,\perp}^{Df}(\chi) A_{\epsilon} B_{\rm ext}^{2}
\label{ndf}
\end{equation}
with
\begin{equation}
I_{\parallel,\perp}^{Df}(\chi)=
\left\{ \begin{array}{ll}
\left[\left(7\right)_{\parallel},\left(4\right)_{\perp}\right] & \textrm { for  } \chi \ll 1 \\
-\frac{9}{7}\frac{45}{2}\frac{\pi^{1/2}2^{1/3}\left(\Gamma\left(\frac{2}{3}\right)\right)^{2}}{\Gamma\left(\frac{1}{6}\right)}\chi^{-4/3}\left[\left(3\right)_{\parallel},\left(2\right)_{\perp}\right] & \textrm{ for   }\chi\gg 1
\end{array}\right.\nonumber
\label{idf}
 \end{equation}
and
\begin{equation}
A_{\epsilon}=\frac{2}{45\mu_{0}}\frac{\epsilon^{4}\alpha^{2} \mathchar'26\mkern-10mu\lambda_\epsilon^{3}}{m_{\epsilon}c^{2}}
\end{equation}
in analogy to equation (\ref{Ae}).
In the limit of large masses ($\chi\ll1$) this expression reduces to (\ref{index}) with the substitution of $\epsilon e$ with $e$ and $m_{\epsilon}$ with $m_{e}$ in equation (\ref{ndf}). The dependence on $B_{\rm ext}$ remains the same as for the well known QED prediction.

For small masses ($\chi\gg1$) the index of refraction now also depends on the parameter $\chi^{-4/3}$ resulting in a net dependence of $n$ with $B_{\rm ext}^{2/3}$ rather than $B_{\rm ext}^{2}$. 

In both mass regimes, a birefringence is induced:
\begin{eqnarray}
&\Delta n^{Df}&=\left[I_{\parallel}^{Df}(\chi)-I_{\perp}^{Df}(\chi)\right] A_{\epsilon} B_{\rm ext}^{2}=\\
&=&\left\{\begin{array}{ll}
3 A_{\epsilon} B_{\rm ext}^{2}& \textrm{ for  } \chi \ll 1 \\
-\frac{9}{7}\frac{45}{2}\frac{\pi^{1/2}2^{1/3}\left(\Gamma\left(\frac{2}{3}\right)\right)^{2}}{\Gamma\left(\frac{1}{6}\right)}\chi^{-4/3}A_{\epsilon} B_{\rm ext}^{2}& \textrm{ for   }\chi\gg 1
\end{array}\right.\nonumber
\label{deltandf}
\end{eqnarray}

 - Spin-0 charged bosons
 
 Very similar expressions to the Dirac fermion case can also be obtained for the spin-0 (s0) charged particle case.\cite{Gies,spin0} Again there are two mass regimes defined by the same parameter $\chi$ of expression (\ref{chi}). In this case the indices of refraction for the two polarization states with respect to the magnetic field direction are
 \begin{equation}
n_{\parallel,\perp}^{s0}=1+I_{\parallel,\perp}^{s0}(\chi) A_{\epsilon} B_{\rm ext}^{2}
\label{ns0}
 \end{equation}
 with
 \begin{equation}
I_{\parallel,\perp}^{s0}(\chi)
=\left\{ \begin{array}{ll}
\left[\left(\frac{1}{4}\right)_{\parallel},\left(\frac{7}{4}\right)_{\perp}\right] & \textrm { for  } \chi \ll 1 \\
-\frac{9}{14}\frac{45}{2}\frac{\pi^{1/2}2^{1/3}\left(\Gamma\left(\frac{2}{3}\right)\right)^{2}}{\Gamma\left(\frac{1}{6}\right)}\chi^{-4/3}\left[\left(\frac{1}{2}\right)_{\parallel},\left(\frac{3}{2}\right)_{\perp}\right] & \textrm{ for   }\chi\gg 1
\end{array}\right.\nonumber
\label{is0}
 \end{equation}
 
The vacuum magnetic birefringence is therefore
\begin{eqnarray}
&\Delta n^{s0}&=\left[I_{\parallel}^{s0}(\chi)-I_{\perp}^{s0}(\chi)\right] A_{\epsilon} B_{\rm ext}^{2}=\\
&=&\left\{\begin{array}{ll}
-\frac{6}{4} A_{\epsilon} B_{\rm ext}^{2}& \textrm{ for  } \chi \ll 1 \\
\frac{9}{14}\frac{45}{2}\frac{\pi^{1/2}2^{1/3}\left(\Gamma\left(\frac{2}{3}\right)\right)^{2}}{\Gamma\left(\frac{1}{6}\right)}\chi^{-4/3}A_{\epsilon} B_{\rm ext}^{2}& \textrm{ for   }\chi\gg 1
\end{array}\right.\nonumber
\label{deltans0}
\end{eqnarray} 
As can be seen there is a sign difference in the birefringence $\Delta n$ induced by an external magnetic field in the presence of Dirac fermions with respect to the case in which spin-0 particles exist. 
 \subsection{Higher order QED corrections}
Figures \ref{4photon} c) and d) show the Feynman diagrams for the $\alpha^{3}$ contribution to the vacuum magnetic birefringence. The effective Lagrangian density for this correction has been evaluated by different authors\cite{rad}\cdash\cite{dunne} and can be expressed as
\begin{equation}
{\cal L}_{\rm Rad} = \frac{A_{e}}{\mu_{0}}\left(\frac{\alpha}{\pi}\right)\frac{10}{72}\left[32\Big(\frac{E^{2}}{c^{2}}-B^{2}\Big)^{2}+263\Big(\frac{\vec{E}}{c}\cdot\vec{B}\Big)^{2}\right]
\end{equation}
This Lagrangian leads to an extra correction $\Delta n_{\rm Rad}$ of the vacuum magnetic birefringence given in equation (\ref{birifQED})
\begin{equation}
\Delta n_{\rm Rad}=\frac{25\alpha}{4\pi}3 A_{e}B_{\rm ext}^{2}=0.0145 \cdot 3 A_{e}B_{\rm ext}^{2}
\end{equation}
resulting in a 1.45 \% correction.

 \section{Apparatus and Method}
The aim of the PVLAS collaboration is to build an apparatus capable of measuring very small ellipticities and rotations. In particular the ultimate goal is to measure  the vacuum magnetic birefringence predicted by the Euler-Heisenberg Lagrangian. At present the best experimental result is an upper bound on $A_e$.\cite{scatter} This bound is a factor of about 5000 above the predicted value. A sensitive ellipsometer designed to measure small birefringences can also be used to measure small rotations due to dichroism. Although QED does not predict dichroism (photon splitting is unmeasurably small), ALP's and MCP's could. 

As will be shown below a birefringence $\Delta n$ will induce an ellipticity $\Psi$ on a linearly polarized beam of light given by
\begin{equation}
\Psi = \pi \frac{L_{\rm eff} {\Delta n}}{\lambda}\sin2\vartheta
\end{equation}
where $L_{\rm eff}$ is the effective path length within the birefringent region with birefringence $\Delta n$ and $\lambda$ is the wavelength of the light traversing it. The induced ellipticity also depends on the angle $\vartheta$ between the light polarization and the magnetic field direction. In the QED case $\Delta n$ depends quadratically on the magnetic field $\vec{B}$. Therefore the magnetic field region must be as long as possible, the magnetic field as intense as possible and the wavelength small. Finally the expected ellipticity must be compared to the different noise sources present and to the maximum available integration time.

Experimentally $L_{\rm eff}$ can be made very long by using a very high finesse Fabry-Perot cavity. In fact given a birefringent region of length $L$ within a Fabry-Perot cavity of finesse $\cal F$ the effective path length is $L_{\rm eff} = \frac{2 \cal F}{\pi}L$. Today finesses ${\cal F} > 400000$ can be obtained.

High magnetic fields can be obtained with superconducting magnets but as we will see below it is desirable to have a time dependent field either by ramping it, thereby changing $\Delta n$ or by rotating the field direction, thereby changing $\vartheta$. This makes superconducting magnets far less appealing than permanent magnets which, today, can reach fields above 2.5 T. Furthermore permanent magnets are relatively inexpensive to buy, have no running costs and have 100\% duty cycle allowing in principle very long integration times.

As for the wavelength we are working with a Nd:YAG laser emitting radiation at $1064$ nm. Frequency doubled versions exist and could double the induced ellipticity but at the moment the highest finesses have been obtained without the frequency doubling.

Lastly it is necessary to make the magnetic field time dependent to move away from DC measurements and limit $1/f$ noise. Two detection schemes exist: homodyne detection or heterodyne detection. This second technique has been adopted in the PVLAS collaboration.

\begin{figure}[pb]
\centerline{\psfig{file = 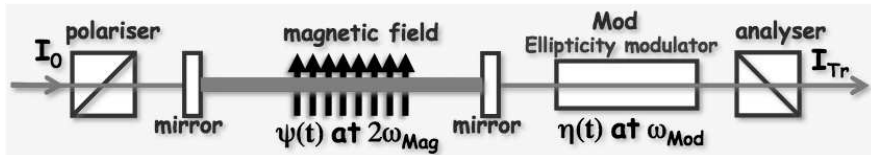, width = 12 cm}}
\caption{Scheme of the PVLAS ellipsometer.}
\label{Scheme}
\end{figure}

A scheme of the ellipsometer is shown in Figure \ref{Scheme}.
The input polarizer linearly polarizes the laser beam of intensity $I_{0}$ which then enters the sensitive region delimited by the Fabry-Perot cavity mirrors. The laser is phase locked to this cavity thus increasing the optical path length within the magnetic field by a factor $2{\cal F}/\pi$ where ${\cal F}$ is the finesse of the cavity. After the cavity the laser beam passes through a photo-elastic ellipticity modulator (PEM) which adds a known time dependent ellipticity $\eta (t)$ to the beam. This modulator ellipticity adds to the ellipticity $\Psi (t)$ acquired within the magnetic field region. After the PEM the beam passes through the analyzer which selects the polarization perpendicular to the input polarization. A photodiode detects $I_{\rm Tr}$ and its Fourier spectrum is then analyzed.

\subsection{Numbers}
To better understand what follows it is useful to present some numerical values of the different quantities involved in the PVLAS experiment. Considering the vacuum magnetic birefringence due to the Euler-Heisenberg Lagrangian, let us determine the ellipticity we expect in the apparatus under construction. We will have a total magnetic field length $L = 1.8$ m with a field intensity $|\vec B_{\rm ext} | = 2.5$ T resulting in $B^{2}_{\rm ext}L = 11.25$ T$^2$m. At present we are running with a finesse ${\cal F} =  240000$. In the past we have reached a maximum finesse value of ${\cal F} = 414000$. Such values have also been published by other authors. We can therefore assume a value of ${\cal F} =  400000$ for our calculation. Putting these numbers together leads to
\begin{equation}
\Psi_{\rm PVLAS} = 2{\cal F}\frac{3 A_{e} L B_{\rm ext}^{2}}{\lambda} = 3.3\cdot 10^{-11}
\end{equation}
Assuming a maximum integration time $T_{\rm max} = 10^{6}$ s and a signal to noise ratio SNR = 1 implies that the sensitivity must be 
\begin{equation}
s_{\rm PVLAS} < \Psi_{\rm PVLAS}\sqrt{T_{\rm max}} = 3.3\cdot 10^{-8} \;\frac{1}{\sqrt{\rm Hz}}
\end{equation}

As discussed in Ref.~\refcite{scatter} the ultimate shot noise ellipticity sensitivity limit with the heterodyne technique depends only on the current generated in the photodiode. Assuming the power output from the cavity $I_{\rm Tr} = 5$~mW and the quantum efficiency of the diode $q = 0.7$ A/W.
\begin{equation}
s_{\rm shot} = \sqrt{\frac{e}{2I_{\rm Tr}q}} \sim 5\cdot10^{-9}\; \frac{1}{\sqrt{\rm Hz}} 
\end{equation}  

At present our sensitivity with $I_{\rm Tr} = 5$ mW is about $s_{\rm Exp} \sim 3\cdot10^{-7}\;\frac{1}{\sqrt{\rm Hz}}$ at the frequency of interest. Work is underway to try to understand the noise present.
\subsection{Heterodyne technique}
\begin{figure}[pb]
\centerline{\psfig{file = 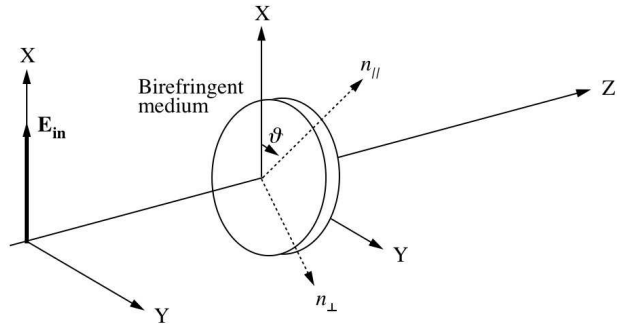, width = 10 cm}}
\caption{Reference frame for the calculations using the Jones matrix formalism. The birefringent medium has a thicness $L$.}
\label{SistRif}
\end{figure}

Considering the coherence of the light source a full treatment of the system can be done with the Jones matrix formalism.\cite{Jones} For the purpose of our discussion let the laser beam propagate along the $Z$ axis and let the incoming (linear) polarization define the $X$ axis (Figure \ref{SistRif}). The Jones matrix for a uniaxial birefringent element of length $L$ is given by
\begin{equation}
\mathbf{BF}(\vartheta)=
\left(
\begin{array}{cc}
1+\imath\psi\cos{2\vartheta} & \imath\psi\sin{2\vartheta} \\
 \imath\psi\sin{2\vartheta} & 1-\imath\psi\cos{2\vartheta}
 \end{array}
 \right)
 \end{equation}
where $\psi$ ($\psi\ll1$) is the induced ellipticity acquired by the light, $\vartheta$ represents the angle between the slow axis ($n_\parallel > n_{\perp}$) of the medium and the $X$ axis. Furthermore 
\begin{equation}
\psi=\frac{\varphi_{\parallel}-\varphi_{\perp}}{2}=\pi\frac{ L(n_{\parallel}-n_{\perp})}{\lambda}
\label{psi}
\end{equation}
with $\varphi_{\parallel}-\varphi_{\perp}$ the phase delay between the parallel and perpendicular polarization components acquired in the length $L$. 

The entrance polarizer defines the input electric field $\vec{E}_{\rm in}=E_{0}{1 \choose 0}$ which, after the magnetic field region, will be
\begin{eqnarray}
\nonumber
\vec{E}_{\rm 0}&=&E_{0}\mathbf{\cdot BF}\cdot{1 \choose 0} = E_{0}
\left(
\begin{array}{c}
1+\imath\psi\cos{2\vartheta} \\
\imath\psi\sin{2\vartheta} 
\end{array}
\right)
\end{eqnarray}
Assuming no losses, the power $I_{\rm Tr}$ after the analyzer (polarizer crossed with respect to the entrance polarizer) will therefore be
\begin{equation}
I_{\rm Tr}=I_{\rm 0}\left|\imath\psi\sin{2\vartheta}\right|^{2}
\end{equation}

The output power is proportional to $\psi^2$ and given the predicted value results in an unmeasurably small intensity component.

By adding a known sinusoidal ellipticity $\eta(t)$ generated with the PEM, the ellipticity signal $\psi$ is linearized. In fact the Jones matrix for the modulator is the same as $\mathbf{BF}$ with $\vartheta$ set at an angle of $\pi/4$ ($\psi\ll\eta\ll1$):
\begin{equation}
\mathbf{MOD}=
\left(
\begin{array}{cc}
1& \imath\eta(t) \\
 \imath\eta(t) & 1
 \end{array}
 \right)
 \end{equation}
 The resulting vector describing the electric field after the modulator will be 
\begin{equation}
\vec{E}_{\rm out}=E_{\rm 0}\mathbf{\cdot MOD \cdot BF}\cdot{1 \choose 0}
\nonumber=E_{\rm 0}
\left(
\begin{array}{c}
1+\imath\psi\cos{2\vartheta}-\psi\eta(t)\sin2\vartheta \\
\imath\eta(t)+\imath\psi\sin{2\vartheta}-\eta(t)\psi\cos2\vartheta
\end{array}
\right)
\end{equation}
Neglecting second order terms, the power $I_{\rm Tr}$ after the analyzer will be
\begin{equation}
I_{\rm Tr}(t)=I_{\rm 0}\left|\imath\eta(t)+\imath\psi\sin{2\vartheta}\right|^{2} \simeq I_{\rm 0}\left[\eta(t)^2+2\eta(t)\psi\sin2\vartheta\right]
\end{equation}
which now depends linearly on the ellipticity $\psi$. To complete the discussion, one finds experimentally that static and slowly varying ellipticities, indicated as $\alpha(t)$, are always present in an real apparatus and that two crossed polarizers have an intrinsic extinction ratio $\sigma^2$, mainly due to imperfections in the calcite crystals. Furthermore, losses in the system reduce the total light reaching the analyzer which we will now indicate as $I_{\rm out}$. Therefore, taking into account an additional 
spurious ellipticity term $\alpha(t)$ (since $\alpha, \psi, \eta \ll 1$ these terms commute and therefore add up algebraically) and a term proportional to $\sigma^2$, the total power at the output of the analyzer will be
\begin{eqnarray}
\nonumber
I_{\rm Tr}(t)&=&I_{\rm out}\left[\sigma^2+\left|\imath\eta(t)+\imath\psi\sin{2\vartheta}+\imath\alpha(t)\right|^{2}\right]\simeq\\
&\simeq&I_{\rm out}\left[\sigma^2+\eta(t)^2+\alpha(t)^{2}+2\eta(t)\psi\sin2\vartheta+2\eta(t)\alpha(t)\right]
\end{eqnarray}

To be able to distinguish the large term $\eta(t)\alpha(t)$ from the term $\eta(t)\psi\sin2\vartheta$, $\psi\sin2\vartheta$ is also modulated in time. This can be done by either ramping the magnetic field intensity (varying therefore $\psi$) or by rotating the magnetic field direction (varying $\vartheta$). The final expression, explicitly indicating the time dependence of $\psi$ and $\vartheta$, for the power at the output of the analyzer is therefore
\begin{equation}
I_{\rm Tr}(t)=I_{\rm out}\left[\sigma^2+\eta(t)^2+\alpha(t)^2+2\eta(t)\psi(t)\sin2\vartheta(t)+2\eta(t)\alpha(t)\right]
\label{spurious}
\end{equation}
\subsection{Optical path multiplier}
To increase the ellipticity induced by the birefringent region of length $L$ one can increase the number of passes through it. Either a multi-pass cavity or a Fabry-Perot cavity can be used for this purpose. In the PVLAS experiment described below, a Fabry-Perot has been chosen.

In a multi-pass cavity the induced ellipticity is proportional to the number of passes $N_{\rm pass}$ through the region. With a Fabry-Perot cavity the analogy to a multi-pass cavity is 
 not immediate since one is dealing with a standing wave.

Let $t$, $r$ be the transmission and reflection coefficients, and $p$ the losses of the mirrors of the cavity such that $t^2+r^2+p=1$. Let $d$ be the length of the cavity and $\delta={4\pi d}/{\lambda}$ the roundtrip phase for a beam of wavelength $\lambda$. Then the Jones matrix for the elements of the ellipsometer after the entrance polarizer is
\begin{equation}
\mathbf{ELL}=
\mathbf{A}\cdot\mathbf{SP}\cdot\mathbf{MOD}\cdot{\it t}^{2}e^{\imath\delta/2}\sum_{n=0}^{\infty} {\left[\mathbf{BF}^{2}{\it r}^{2}e^{\imath\delta}\right]^{n}}\cdot\mathbf{BF}
 \end{equation} 
where $\mathbf{A}=\left(
\begin{array}{cc}
0 & 0 \\
0 & 1
 \end{array}
 \right)$ is the analyzer Jones matrix and $\mathbf{SP}$ describes the spurious ellipticity mainly due to the mirrors of the cavity itself.
 Because ${\it r}^{2} < 1$, $\mathbf{ELL}$ can be rewritten as
\begin{equation}
\mathbf{ELL}=
\mathbf{A}\cdot\mathbf{SP}\cdot\mathbf{MOD}\cdot{\it t}^{2}e^{\imath\delta/2}{\left[\mathbf{I}-\mathbf{BF}^{2}{\it r}^{2}e^{\imath\delta}\right]^{-1}}\cdot\mathbf{BF}
\label{fp}
\end{equation} 
with $\mathbf{I}$ the identity matrix. With the laser phase locked to the cavity so that $\delta=2\pi m$, where $m$ is an integer number, 
the electric field at the output of the system will be
\begin{equation}
\vec{E}_{\rm out}=E_{\rm 0}\mathbf{\cdot ELL}\cdot{1 \choose 0}
=E_{\rm 0}\frac{{\it t}^{2}}{{\it t}^{2}+p}
\left(
\begin{array}{c}
0 \\
\imath\alpha(t)+\imath\eta(t)+\imath\frac{1+{\it r}^{2}}{1-{\it r}^{2}}\psi\sin{2\vartheta}
\end{array}
\right)
\end{equation}
and the power, including losses,
\begin{equation}
I_{\rm Tr}(t)=I_{\rm out}\Bigg|\imath\alpha(t)+\imath\eta(t)+\imath\left(\frac{1+{\it r}^{2}}{1-{\it r}^{2}}\right)\psi\sin{2\vartheta}\Bigg|^{2}
\label{Iout}
\end{equation}
This expression is at the basis of the ellipsometer in the PVLAS apparatus.\cite{dicindotto}
Small ellipticities add up algebraically and the Fabry-Perot multiplies the single pass ellipticity $\psi\sin{2\vartheta}$, generated within the cavity, by a factor $({1+{\it r}^{2}})/({1-{\it r}^{2}})\approx{2{\cal F}}/{\pi}$, where ${\cal F}$ is the finesse of the cavity. The ellipticity signal to be detected is therefore $\Psi = ({2{\cal F}}/{\pi})\psi\sin{2\vartheta}$.
\subsection{Fourier components}
In the PVLAS experiment, $\eta(t)=\eta_{0}\cos(\omega_{\rm Mod}t+\theta_{\rm Mod})$ and the magnetic field direction is rotated at an angular velocity $\Omega_{\rm Mag}$. A Fourier analysis of the power $I_{\rm Tr}(t)$ of equation (\ref{Iout}) results in four main frequency components each with a definite amplitude and phase. These are reported in table \ref{components}.
\begin{table}[h!]
\caption{Intensity of the frequency components of the signal after the analyzer $\mathbf{A}$.}
\begin{tabular}{c|c|c|c}
\hline\noalign{\smallskip}
Frequency & Fourier component & Intensity/$I_{0}$ & Phase\\
\noalign{\smallskip}\hline\noalign{\smallskip}
$DC$ & $I_{\rm DC}$ & $\sigma^2+\alpha_{\rm DC}^2+\eta_{0}^{2}/2$ & $-$\\
$\omega_{\rm Mod}$ & $I_{\omega_{\rm Mod}}$ & $2\alpha_{\rm DC}\eta_{0}$ & $\theta_{\rm Mod}$\\
$\omega_{\rm Mod}\pm2\Omega_{\rm Mag}$ & $I_{\omega_{\rm Mod}\pm2\Omega_{\rm Mag}}$ & $\eta_{0}\frac{2{\cal F}}{\pi}\psi$ & $\theta_{\rm Mod}\pm2\theta_{\rm Mag}$\\
$2\omega_{\rm Mod}$ & $I_{2\omega_{\rm Mod}}$ & $\eta_{0}^{2}/2$ & $2\theta_{\rm Mod}$\\
\noalign{\smallskip}\hline
\end{tabular}
\label{components}
\end{table}

The presence of a component at $\omega_{\rm Mod}\pm2\Omega_{\rm Mag}$ in the signal identifies an induced ellipticity within the Fabry-Perot cavity. Furthermore the phase of this component must satisfy the value in table \ref{components}.
\section{Experimental studies}
\subsection{PVLAS - LNL limitations}
Although the previous PVLAS apparatus\cite{scatter} setup at LNL in Legnaro, Italy, set best limits on magnetic vacuum birefringence and photon-photon elastic scattering at low energies several limitations were present in this apparatus:

\begin{itemize}
\item stray field due to the superconducting magnet when operating at high fields
\item limited running time due to liquid helium consumption in the rotating cryostat
\item high running costs
\item seismic noise
\item with a single magnet a zero measurement with the same experimental conditions as with the field ON is not possible 
\end{itemize}

To solve the first two points in the new setup we chose to work with permanent magnets instead of the superconducting magnet. In this way the stray field will be much smaller and the duty cycle will be 100\%. Furthermore studies on the PVLAS - LNL setup revealed limitations due to seismic noise. Due to the size and configuration of the optical benches this problem could not be solved directly on site. As was shown in \refcite{towardsQED} it is necessary to seismically isolate the whole ellipsometer apparatus on a single optical bench.

For these reasons a complete new setup is being rebuilt on a single granite optical bench.

\subsection{Two magnet configuration}
The last point in the above list deserves some attention. A zero measurement in an experimental condition as close to the signal configuration as possible is absolutely mandatory after the experience with the PVLAS - LNL apparatus. Especially when dealing with such a sensitive apparatus. Therefore instead of using a single dipole magnet, two magnets will be used. By orienting the fields of the two magnets at $90^{\circ}$ and assuming the magnets to be identical the net ellipticity generated by the magnetic birefringence is zero. 
Running the system with the magnets parallel and perpendicular will allow the identification of a real physical signal with respect to some spurious signal due to stray field.

\begin{figure}.
\centerline{\psfig{file = 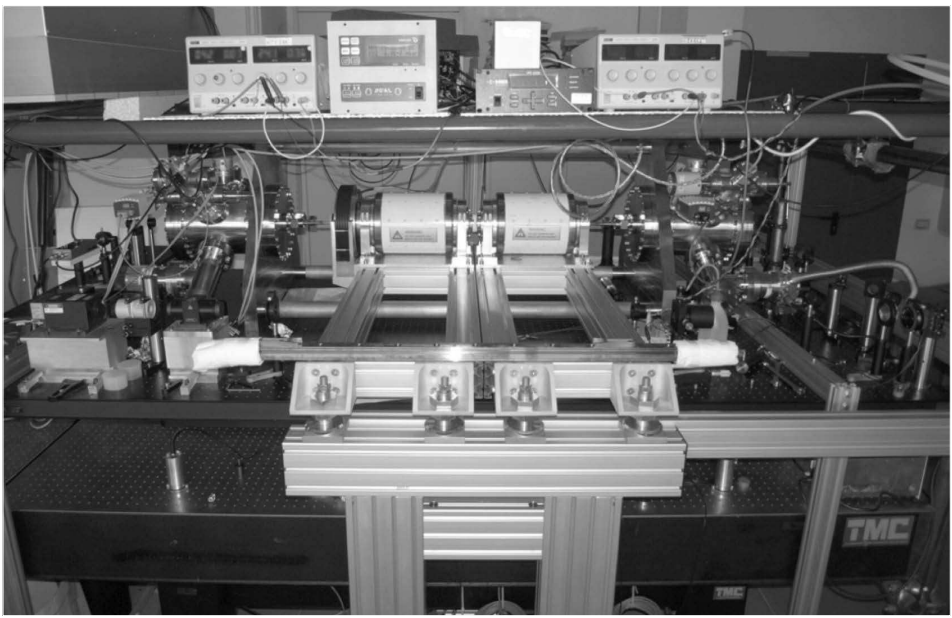, width = 13 cm}}
\caption{Photograph of the test apparatus in Ferrara. At the center one can see the two dipole permanent magnets. The optics is supported by two antivibration stages whereas the magnet supports are on the floor.}
\label{testsetup}
\end{figure}

\begin{figure}[pb].
\centerline{\psfig{file = 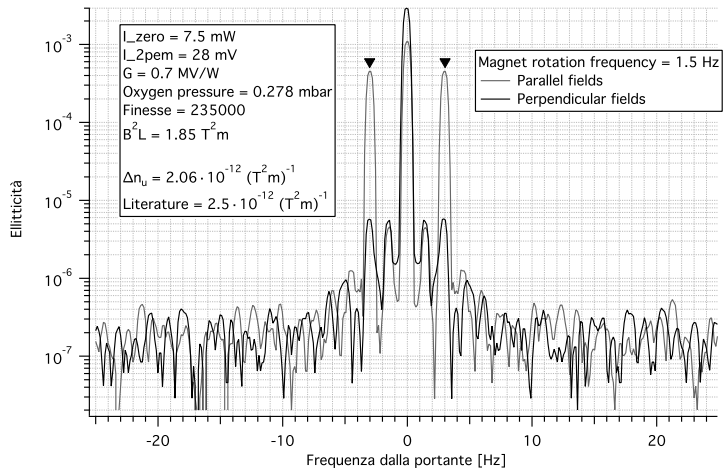, width = 10 cm}}
\caption{Fourier spectrum around the carrier frequency $\omega_{\rm Mod}$. Ellipticity measurements with the magnets parallel (grey curve) and the magnets perpendicular (black curve). The large sidebands at $2\Omega_{\rm Mag}$ are due to the Cotton Mouton effect in Oxygen gas.}
\label{CM_O2}
\end{figure}

A verification of this idea was done with our test setup in Ferrara using the Cotton-Mouton effect of oxygen. The principle of the test ellipsometer is identical to the final system in construction. A photograph of the test apparatus is shown in Figure \ref{testsetup}. At the center one can see the two permanent magnets each generating a 20 cm long magnetic field of intensity ${|\vec B}| = 2.3 $~T. The whole optical setup is placed on a seismically isolated optical bench whereas the magnets are supported by a structure placed on the floor thereby mechanically isolated from the optics.
Figure \ref{CM_O2} shows the Fourier spectrum around the carrier frequency $\omega_{\rm Mod}$. Clear sidebands can be seen at twice the rotation frequency of the magnets ($\Omega_{\rm Mag} = 1.5$~Hz). Shown in grey is the Fourier transform with the magnets in a parallel configuration whereas in black the magnets are perpendicular. The signal attenuation factor in the perpendicular configuration with respect to the parallel one is about 80. Although we also see sidebands at once the magnet rotation frequency (which should in principle not be there) which means we may have a small spurious component even a $2\Omega_{\rm Mag}$, we can conclude that the parameter $B_{\rm ext}^{2}L$ for the two magnets is equal to within about 1-2\% and most importantly that the principle is correct.
We believe this improvement will be crucial in understanding the ellipsometer and that a measurement with two magnets whose field directions can be changed is imperative. There is no way of guaranteeing the authenticity of a signal observed with only one magnet or with several magnets all in a parallel configuration.
\subsection{Preliminary measurements}
With the Ferrara test setup (Figure \ref{testsetup}) measurements have been performed to understand its limits and optimize the new apparatus in construction.
Two different noise sources exist and are under study: wideband noise and signals at the magnet rotation frequency and its harmonics. Below we briefly report some results in the two cases. 
\subsubsection{Sensitivity - wideband noise}
Measurements were first performed without the Fabry-Perot cavity. We successfully exclude noise sources from readout electronics and optical elements other than the cavity mirrors reaching the expected sensitivity of $s_{\rm no\;cavity} = 6\cdot10^{-9}\;\frac{1}{\sqrt{\rm Hz}}$ and a noise floor of $\psi_{\rm floor} = 1-2\cdot10^{-10}$ with $1600$ s integration time.

With the introduction of the cavity with finesse ${\cal F} = 240000$ the noise increased to $s_{\rm cavity} = 3\cdot 10^{-7}\;\frac{1}{\sqrt{\rm Hz}}$ at about 6 Hz. This was significantly more than what was expected from the reduction of $I_{\rm out}$ due to cavity losses. This unexplained noise is under study and we suspect variations of the intrinsic birefringence of the mirrors.

We also showed that the magnet rotation did not contribute to the wideband noise indicating a good isolation between the magnet support and optical setup.
\subsubsection{Spurious peaks}
With the magnets in rotation we often observe ellipticity peaks varying from a few $10^{-8}$ to a few $10^{-7}$ whereas sometimes such peaks are not present. The frequencies of these peaks are at harmonics of the magnet rotation frequency. The variability of these peaks from one run to another seems to depend (in a non reproducible way) on the adjustment of the input and output polarizers which are done with motorized stages. To study the dependence of such peaks on the magnet orientation, a field probe is present near the output side magnet. Changing the relative orientation of the two magnets does not change the amplitude of these peaks but does change their phase indicating that the more sensitive part of the apparatus seems to be the entrance optics. All the motorized stages have small electric motors which couple to the rotating magnetic field and may introduce beam jitter and therefore ellipticity. The substitution of all these stages is programmed shortly. 
\subsubsection{Noise floor measurements}
\begin{figure}.
\centerline{\psfig{file = 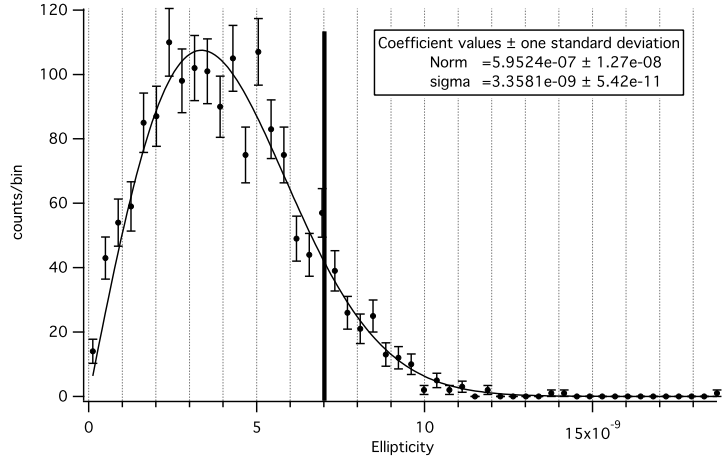, width = 10 cm}}
\caption{Histogram of ellipticity noise in a frequency band around $2\Omega_{\rm Mag}$: $(3.25\pm0.39)$~Hz. The integration time was $T = 8192$~s. Superimposed is a fit with the Rayleigh distribution resulting in an ellipticity standard deviation $\sigma = 3.36\cdot 10^{-9}$. The vertical black line indicates the value in the ellipticity Fourier spectrum bin corresponding to $2\Omega_{\rm Mag} = 3.25$~Hz.}
\label{limite}
\end{figure}
With the apparatus in a condition in which the peak a $2\Omega_{\rm Mag}$ is not present measurements of a few hours have been done. The magnet rotation frequency was $\Omega_{\rm Mag} = 1.625$~Hz. In Figure \ref{limite} we report the histogram of the ellipticity noise in a narrow band around $2\Omega_{\rm Mag}$: $(3.25 \pm 0.39)$ Hz. The integration time in this example was $T = 8192$ s. The probability density function for a noise amplitude with equal standard deviations $\sigma$ for the 'in phase' and quadrature components is the Rayleigh function: $P(r) = \frac{r}{\sigma^{2}}e^{-\frac{r^2}{2\sigma^2}}$. In Figure \ref{limite} we have superimposed a fit with the Rayleigh function in which the ellipticity standard deviation is $\sigma = 3.36\cdot10^{-9}$. A vertical line in the same figure indicates the value in the Fourier spectrum bin corresponding to exactly $2 \Omega_{\rm Mag} = 3.25$~Hz.
Given the finesse ${\cal F} = 240000$, $L = 0.4 $~m, $|\vec{B}| = 2.3$~T and $\lambda = 1064$~nm the value $\sigma = 3.36\cdot10^{-9}$ translates, at 95 \% c.l., in a birefringence limit induced by the magnetic field of
\begin{equation}
\Delta n < \frac{\sigma\lambda}{2{\cal F}L} = 4.6\cdot10^{-20}
\end{equation}
The parameter $A_e$ and the light-light elastic scattering cross section limit can also be deduced (at 95 \% c.l.):
\begin{eqnarray}
A_e& <& \frac{\Delta n}{3 B_{\rm ext}^{2}} = \frac{\sigma\lambda}{2{\cal F}L3 B_{\rm ext}^{2}} = 2.9\cdot10^{-21}\;{\rm T}^{-2}\\
\sigma_{\gamma\gamma}& <& 9.5\cdot10^{-59}\;{\rm cm}^{2}\; @ \;1064\;{\rm nm}
\end{eqnarray}
Although not always reproducible, these limits are about a factor 2 better than the best previously published limits obtained with the PVLAS - LNL apparatus.\cite{scatter}
\section{Conclusions}
We have presented the physics the PVLAS experiment is aiming at studying and have briefly discussed the experimental method. Noise sources are being studied on a bench-top small apparatus in Ferrara, Italy, in view of the construction of the final apparatus with which we hope to measure for the first time the magnetic birefringence of vacuum due to vacuum fluctuations.

We have discussed the importance of using two dipole magnets instead of only one whose directions can be made perpendicular to each other in order to have a zero effect condition with the magnetic field present. This is vital to study and eliminate spurious signals generated by the rotating field. A proof of principle measurement using the Cotton-Mouton effect in Oxygen gas was done with the test apparatus in Ferrara, Italy.

Finally noise floor measurements were performed in conditions in which spurious peaks were not present. A new limit on the parameter $A_e$ describing non linear electrodynamic effects in vacuum has been obtained: $A_e < 2.9\cdot10^{-21}$ T$^{-2}$. This value improves the previous one by a factor 2.2.



\begin{thebibliography}{00}
\bibitem{HypIn}
D. Bakalov {\it et al.}, Hyperfine Interactions {\bf114} (1998) 103.\\
D. Bakalov {\it et al.}, Quantum Semicl. Opt. {\bf 10} (1998) 239.
\bibitem{PRD}
E. Zavattini {\it et al.}, Phys. Rev. D {\bf77} (2008) 032006.
\bibitem{Brodin}
E. Lundstr\"om {\it et al.}, Phys. Rev. Lett. {\bf 96} (2006) 083602.
\bibitem{exawatt}
D. Tommasini {\it et al.}, Phys. Rev. A {\bf 77} (2008) 042101.
\bibitem{Luiten1}
A. N. Luiten and J. C. Petersen, Phys. Lett. A {\bf330} (2004) 429.
\bibitem{Luiten2}
A. N. Luiten and J. C. Petersen, Phys. Rev. A {\bf70} (2004) 033801.
\bibitem{bmv}
R. Battesti {\it et al.}, Eur. Phys. J. D {\bf46}, (2008) 323.
\bibitem{ni}
W.-T. Ni, Chinese J. Phys. {\bf34} (1996) 962.
\bibitem{pugnat}
P. Pugnat {\it et al.}, Czech. J. Phys. A {\bf56} (2006) C193.
\bibitem{heinzl}
T. Heinzl {\it et al.}, Optics Comm. {\bf267} (2006) 318.
\bibitem{QED}
H. Euler and B. Kochel, Naturwiss. {\bf 23} (1935) 246.\\
W. Heisenberg and H. Euler, Z. Phys. {\bf98} (1936) 718.\\
V. S. Weisskopf, Kgl. Danske Vid. Sels., Math.-fys. Medd. {\bf14} (1936) 6.\\
J. Schwinger, Phys. Rev. {\bf82} (1951) 664.
\bibitem{Adler}
R. Baier and P. Breitenlohner, Acta Phys. Austriaca {\bf25} (1967) 212.\\
R. Baier and P. Breitenlohner, Nuovo Cimento {\bf47} (1967) 261.\\
S. L. Adler, Ann. Phys. {\bf67} (1971) 559.\\
Z. Bialynicka-Birula and I. Bialynicki-Birula, Phys. Rev. D {\bf2} (1970), 2341.
\bibitem{Iacopini}
E. Iacopini and E. Zavattini, Phys. Lett. B {\bf85} (1979) 151.
\bibitem{DeTollis}B. De Tollis, Nuovo Cimento {\bf35} (1965) 1182.\\
B. De Tollis, Nuovo Cimento {\bf32}, (1964) 757.
\bibitem{Karplus}
R. Karplus {\it et al.}, Phys. Rev. {\bf83} (1951) 776.
\bibitem{Duane}
D. A. Dicus {\it et al.}, Phys. Rev. D {\bf57} (1998) 2443.
\bibitem{Bernard}
D. Bernard {\it et al.}, Eur. Phys. J. D {\bf10} (2000) 141.
\bibitem{BernardOld}
F. Moulin {\it et al.}, Z. Phys. C {\bf72} (1996) 607.
\bibitem{Haissinsky}
J. Ha\"issinski {\it et al.} Phys. Scr. {\bf74} (2006) 678-681.
\bibitem{Denisov}
V. I. Denisov {\it et al.}, Phys. Rev. D {\bf69} (2004) 066008.
\bibitem{Petronzio}
L. Maiani  {\it et al.}, Phys. Lett. B{\bf 175} (1986) 359.
\bibitem{Sikivie}
P. Sikivie, Phys. Rev. Lett. {\bf 51} (1983) 1415.
\bibitem{Gasperini}
M. Gasperini, Phys. Rev. Lett. {\bf 59} (1987) 396. 
\bibitem{Raffelt}
G. Raffelt and L. Stodolsky, Phys. Rev. D {\bf37} (1998) 1237.
\bibitem{Gies}
M. Ahlers {\it et al.}, Phys. Rev. D {\bf75} (2007) 035011.
\bibitem{Ringwald}
H. Gies {\it et al.}, Phys. Rev. Lett. {\bf97} (2006) 140402.
\bibitem{fermion}
W. y. Tsai and T. Erber, Phys. Rev. D {\bf12} (1975) 1132.
\bibitem{landaulev}
J. K. Daugherty {\it et al.}, Astrophys. J. {\bf273} (1983) 761.
\bibitem{spin0}
C. Schubert, Nucl. Phys. {\bf B585} (2000) 407.
\bibitem{qqbar1}
M. Davier et al., arXiv:0908.4300[hep-ph] and refs. therein
\bibitem{qqbar2}
A. Nyffeler, arXiv:1001.3970[hep-ph] and refs. therein
\bibitem{scatter}
M. Bregant {\it et al.}, Phys Rev. D {\bf 78} (2008) 032006.
\bibitem{towardsQED}
F. Della Valle {\it et al.}, Optics Comm. {\bf 293} (2010) 4194.
\bibitem{Cameron}
R. Cameron {\it et al.}, Phys. Rev. D {\bf47} (1993) 3707.
\bibitem{PRL}
E. Zavattini {\it et al.}, Phys. Rev. Lett. {\bf96} (2006) 110406.
\bibitem{CAST}
K. Zioutas {\it et al.}, Phys. Rev. Lett. {\bf94} (2005) 121301.
\bibitem{RizzoALP1}
C. Robilliard {\it et al.}, Phys. Rev. Lett. {\bf99} (2007) 190403.
\bibitem{RizzoALP2}
M. Fouch\'e {\it et al.}, Phys. Rev. D {\bf78} (2008) 032013.
\bibitem{gammeV}
A. S. Chou {\it et al.}, arXiv:0710.3783\\
A. S. Chou {\it et al.}, Phys. Rev. Lett. {\bf 100} (2008) 080402.
\bibitem{rad}
V. I. Ritus, Zh. Eksp. Teor. Fiz {\bf69} (1975) 1517 [Sov. Phys. JETP {\bf42}
(1975) 774].
\bibitem{bakalovrad}
D. Bakalov, INFN/AE-94/27 (1994) SIS publication LNF
\bibitem{dunne}
G. Dunne, arxiv:hep-th/0406216v1
\bibitem{Jones}
R. C. Jones, J. Opt. Soc. Am. 38 (1948) 671.\\
E. Hecht, {\it Optics}, 2nd ed., Addison-Wesley, San Francisco (1987).
\bibitem{dicindotto}
G. Zavattini {\it et al.}, Applied Physics B {\bf83} (2006) 571.
\end{thebibliography}
\end{document}